\documentclass[onecolumn,showpacs]{revtex4}
\usepackage{graphicx}
\usepackage{dcolumn}
\usepackage{bm}
\usepackage{amsmath}
\usepackage{amssymb}
\usepackage{multirow}
\usepackage{caption}
\usepackage{epstopdf}
\begin{document}
{\setlength{\oddsidemargin}{1.2in}
\setlength{\evensidemargin}{1.2in} } \baselineskip 0.55cm
\begin{center}
{\LARGE {\bf Qualitative Stability Analysis of Cosmological Parameters in $f(T,B)$ Gravity}}
\end{center}
\date{\today}
\begin{center}
  Amit Samaddar, S. Surendra Singh \\
   Department of Mathematics, National Institute of Technology Manipur,\\ Imphal-795004,India\\
   Email:{ samaddaramit4@gmail.com, ssuren.mu@gmail.com }\\
 \end{center}

 \begin{center}
 \textbf{Abstract}
  \end{center}
We analyze the cosmological solutions of $f(T,B)$ gravity using dynamical system analysis where $T$ is the torsion scalar and $B$ be the boundary term scalar. In our work, we assume two specific cosmological models. For first model, we consider $ f(T,B)=f_{0}(B^{k}+T^{m})$, where $k$ and $m$ are constants. For second model, we consider $f(T,B)=f_{0}T B$.  We generate an autonomous system of differential equations for each models by introducing new dimensionless variables. To solve this system of equations, we use dynamical system analysis. We also investigate the critical points and their natures, stability conditions and their behaviors of Universe expansion. For both models, we get four critical points. The phase plots of this system are analyzed in detail and study their geometrical interpretations also. In both model, we evaluated density parameters such as $\Omega_{r}$, $\Omega_{m}$, $\Omega_{\Lambda}$ and $\omega_{eff}$ and deceleration parameter $(q)$ and find their suitable range of the parameter $\lambda$ for stability. For first model, we get $\omega_{eff}=-0.833,-0.166$ and for second model, we get  $\omega_{eff}=-\frac{1}{3}$. This shows that both the models are in quintessence phase.  Further, we compare the values of EoS parameter and deceleration parameter with the observational values.
  \begin{center}
  \textbf{1.  Introduction}
   \end{center}
    Recent cosmologists observed that our present Universe is expanding with late time acceleration \cite{1}-\cite{2}. To understand that the Universe is accelerated or decelerated, cosmologists introduced a dimensionless variable, which is known as deceleration parameter as $q=-\frac{a\ddot{a}}{\dot{a}^{2}}$. If $q<0$, then the expansion of Universe is accelerated. If $q>0$, then the Universe is expanding with deceleration. The proof towards the accelerating expansion of Universe given by magnitude-redshift relation. In flat homogeneous and isotropic Universe, about $70\%$ energy density consists of some weird components. This weird component of energy is known as dark energy. The value of matter energy density parameter is $\Omega_{m}=0.28^{+0.09}_{-0.08}$. Many cosmological hypothesis are already made to understand correctly the fundamental behavior of dark energy.  Lack of a consistent theory of quantum gravity, the dark energy problem is still unknown, one of the notable nominees is cosmological constant, which gives a negative pressure of with the EoS parameter, $\omega=-1$ \cite{3}. This dark energy was responsible for cosmological acceleration of Universe's expansion. Theoretical models to explain the cosmological acceleration were serialized in two types. The 1st model was Modified Gravity Model. In this model, the General Theory of Relativity(GR) was modified by  \cite{4}.  The important things of these type of models are that the late time acceleration of the Universe can be achieved without considering any exotic matter component. The second model is dark energy model, which is also called as modified matter models. There were so many dark models for dark energy, which was cosmological constant or vaccum energy, quintessence, quantum, phantom energy, tachyon, fermion etc \cite{5}-\cite{7}. To attain the static Universe Einstein founded Cosmological Constant($\Lambda$) in his equation in $1917$. After that Hubble introduced the accelerated expansion of the Universe in $1929$ and then Einstein rejected his idea of cosmological constant. After discovering the late-time acceleration, cosmological constant deliberated as most useful and simplest structure of the dark energy.  For explaining the expanding Universe, general relativity theory(GR) changed by cosmological constant($\Lambda$) which is known as $\Lambda$CDM cosmological model \cite{8}. The $\Lambda$CDM (Lambda cold dark matter) model contains three components $\rightarrow$ $(i)$ cosmological constant($\Lambda$) connected with dark energy, $(ii)$ cold dark matter and $(iii)$ ordinary matter. The $\Lambda$CDM cosmological model satisfied the cosmological principal,  which is, on the largest  scale, the Universe was homogeneous and isotropy; "the Universe look the same wherever you are in the Universe and which direction you look." The $\Lambda$CDM model presumed that the features of the Universe was flat, this means curvature was zero. Moreover, this model was emerged on Einstein's General Theory of Relativity(GR). The cosmological constant($\Lambda$) belongs to negative pressure when $p = -\rho c^{2}$, in accordance with the GR, it reasons to accelerated expansion of the  Universe. In the Universe, value of the dark energy i.e. $\Omega_{\Lambda}$ is $0.669\pm 0.038$ based on $2018$ \cite{9}.\\
 Teleparallel Gravity(TG) is dynamically equivalent to GR where curvature is changed by torsion by the teleparallel connections. Modified form of teleparallel gravity is an alternative method to interpret the cosmic acceleration. This theory is also called $f(T)$ gravity. In this gravity theory, we use the Weitzenb$\ddot{o}$ck connection that has torsion but not curvature, to use curvature constructed by Levi-Civita connection of GR. An important feature of this theory was that it contained the $2nd$ order field equations which was easy to compare the $f(R)$ gravity theory with $4th$ order field equations. Moreover, the dynamical equations of Teleparallel Gravity obtained by substituting the Einstein-Hilbert action with the torsion scalar $T$. With this replacement, we created the $\it Teleparallel$ $\it Equivalent$ $\it of$ $\it General$ $\it Relativity$(TEGR), it was different from General Relativity with the Lagrangian boundary term $B$. In this framework of teleparallel gravity, a new modified gravity called $f(T,B)$ gravity theory, it is reduced to  $f(R)$ and $f(T)$ gravity by taking $f(R)$= $f(-T+B)$. The $f(T,B)$  theory was an attracting gravity theory and models for $f(T,B)$ gravity pursued at all levels of phenomenology \cite{10}.\\
    In our present work, we discuss the behaviors of $f(T,B)$ gravity theory with the help of dynamical system analysis of a isotropic and homogeneous Universe using the FLRW metric. In theories of gravity, the major problem is to find out the analytical or numerical solutions because in field equations many nonlinear terms are present, which is difficult to solve and compare to the observational data is not so easy. So we need some other methods to solve these types of nonlinear equations, and which is also useful to analyze the dynamical behavior and also the stability condition. To solve these types of problem, the well known process is Dynamical System Analysis \cite{11}.  \\
In our work the main motive is to discuss the late time acceleration of $f(T,B)$ gravity with its functional form and analyzed the stability behavior of these models. We discussed mostly on the mathematical expressions and solutions of $f(T,B)$ gravity and geometrical interpretations. This paper is systematized as: In part $2$, we discuss the action of $f(T,B)$ gravity model and obtained gravitational field equations. In part $3$, we consider some dimensionless variables with the help of field equations and find the autonomous system of differential equation. In part $4$, we consider two $f(T,B)$ cosmological models with two functional form of $f(T,B)$. In this part, we use the dynamical system analysis to find the numerical solution of $f(T,B)$ gravity theory. The behaviour of energy conditions are also discussed in this section. In this part, we find the Hubble parameter, deceleration parameter and Eos parameter and draw their plots.  The conclusion is given in part $5$.
\begin{center}
   \textbf{2.  $f(T,B)$ Cosmology}
\end{center}
In our work, we discuss the generalization of the TEGR action to an arbitrary function of both the torsion scalar and the boundary term, which gives
  \begin{equation}
    S_{f(T,B)} = \frac{1}{2k^{2}}\int d^{4}x e f(T,B) + \int d^{4}x e \mathcal{L}_{m}
  \end{equation}
  where $ k^{2}=8\Pi G $, $G$ is Newtonian constant, $ \mathcal{L}_{m} $ is the matter Lagrangian Jordan frame and $e$ be the volume element of the metric tensor that is equal to $\sqrt{-g}$. Here we consider a flat homogeneous and isotropic FLRW metric of Cartesian coordinates which is written as

  \begin{equation}
ds^{2}=-dt^{2}+a^{2}(t)(dx^{2}+dy^{2}+dz^{2})
\end{equation}
where $a(t)$ be the scale factor. We take a transformation over $ \hat{f}(T,B)\rightarrow -T+f(T,B) $, which satisfies the diffeomorphic invariance. $ \hat{f} (T,B)\rightarrow -T+f(T,B)$ illustrates the arbitrary Lagrangian over the torsion scalar and boundary term is diffeomorphic invariant. From equation (1), we found the components in diagonal form as follows:
\begin{equation}
e^{a}_{\mu}= diag(1,a(t),a(t),a(t))
\end{equation}
Here we take the expressions of torsion scalar as
 \begin{equation}
T= 6 H^{2}
\end{equation}
and the boundary term is given by
\begin{equation}
  B= 6(3 H^{2}+\dot{H})
\end{equation}
which together given by the Ricci scalar of the FLRW metric.
\begin{equation}
R=-T+B=6(\dot{H}+ 2H^{2})
\end{equation}
where $H=\frac{\dot{a}}{a}$ be the Hubble Parameter. We assumed that our Universe was filled with dust and radiation. Radiation parameters, $\rho_{r}$ and $p_{r}$ are related to each other with the equation of state $p_{r}=\omega_{r}\rho_{r}$ where the constant $\omega_{r}=\frac{1}{3}$. No interactions are anticipated to occur between the matter and radiation  so the continuity equations for radiation and matter are
\begin{equation}
  \dot{\rho_{m}}+3H\rho_{m}=0
\end{equation}
and
\begin{equation}
   \dot{\rho_{r}}+4H\rho_{r}=0.
\end{equation}
By varying the action with respect to the metric, the field equations can be obtained as
\begin{equation}
  -3H^{2}(3f_{B}+2f_{T})+3H\dot{f_{B}}-3\dot{H}f_{B}+\frac{1}{2}f=k^{2}(\rho_{m}+\rho_{r}).
\end{equation}
\begin{equation}
   -(3H^{2}+\dot{H})(3f_{B}+2f_{T})-2H\dot{f_{T}}+\ddot{f_{B}}+\frac{1}{2}f=-k^{2}(\rho_{m}+\frac{4}{3}\rho_{r}).
\end{equation}
where $\rho_{r}$ and $\rho_{m}$ be the radiation and matter density respectively and $f_{T}=\frac{\partial f}{\partial T}$ and $f_{B}=\frac{\partial f}{\partial B}$.
\begin{center}
 \hspace{1cm}
   \textbf{3.  Dynamical Structure of $f(T,B)$ Gravity }
\end{center}
 \hspace{1cm}
In this part, we abide by the procedure to system of dynamical analysis to check the stability behavior $f(T,B)$ gravity models, which can be accomplished by reconstructing the cosmological equations to dynamical system \cite{12}. Now, we introduce the dimensionless variables from equation $(9)$ as follows:
\begin{equation}
  x=\frac{\dot{f_{B}}}{3Hf_{B}}, \hspace{0.5cm}   y=\frac{f}{18 H^{2}f_{B}}, \hspace{0.5cm}   z=-\frac{\dot{H}}{3H^{2}}, \hspace{0.5cm}   \omega=-\frac{k^{2}\rho_{r}}{9H^{2}f_{B}}, \hspace{0.5cm}  s=-\frac{2f_{T}}{3f_{B}}
\end{equation}
where $ k^{2}=1$. Then the various density parameters are,
\begin{equation}
  \Omega_{r}=-\frac{\rho_{r}}{9H^{2}f_{B}}=\omega,  \hspace{0.5cm}  \Omega_m=-\frac{\rho_{m}}{9H^{2}f_{B}}=1-x-y-z-s-\omega
\end{equation}
where $\Omega$ be the parameter which depends upon the other dynamical variables. We derive the following set of autonomous differential equations from equation (11).
\begin{equation}
  x'=3-3x-6y-6z+9\omega-x^{2}+3zs+3xz+\frac{2\dot{f_{T}}}{3Hf_{B}},
\end{equation}
\begin{equation}
  y'=-6z+\lambda+6yz-3xy,
\end{equation}
\begin{equation}
   z'=\lambda-6z^{2},
\end{equation}
\begin{equation}
  \omega'=-4\omega-3x\omega+6z\omega,
\end{equation}
\begin{equation}
  s'=-\frac{2\dot{f_{T}}}{3Hf_{B}}-3xs .
\end{equation}
\hspace{1cm}
Here $(')$ denotes the derivative with respect to $\eta$, where $\eta= log  a$ denotes the logarithmic time with respect to the scale factor $a$ and $\dot{f_{T}}$ be the derivative w.r.to $t$. For this system, we need to define the parameter
\begin{equation}
  \lambda=-\frac{\ddot{H}}{3H^{3}}
\end{equation}
where the parameter $\lambda$ is the functional form of Hubble rate.  We discussed later that when $\lambda= constant$ \cite{13}, some cosmological solutions can be regained. For example, when  $\lambda=0$, it is produced exact de Sitter scalar factor $a(t)=e^{\Lambda t}$ where $\Lambda$= constant or, form a quasi de Sitter scalar $a(t)= e^{H_{\circ}t+H_{1}t^{2}}$, where $H_{\circ}$ and $H_{1}$ are constants. In Ref. [13], when $\lambda=-\frac{9}{2}$, it produced a matter dominated scalar factor, $a(t)\sim 9 t^{\frac{2}{3}}$. Based of these values of $\lambda$, we determine the characteristics of effective equation of state parameter i.e. $\omega_{eff}$. The general form of effective EoS parameter is
\begin{equation}
  \omega_{eff}= -1-\frac{2}{3}\frac{\dot{H}}{H^{2}},
  \end{equation}
  and it can be written in terms of $z$ which is given as follows:
  \begin{equation}
   \omega_{eff}= -1+2z,
\end{equation}\\
But the total form of EoS determines the significance of the cosmological evolution to the fixed points  which are expressed in terms of $\lambda$. For example, in de Sitter or quasi de Sitter phase for $\lambda=0$, we get the effective EoS parameter  $\omega_{eff}= -1$ and for $\lambda=-\frac{9}{2}$, we get $\omega_{eff}=0$.
\begin{center}
 \hspace{1cm}
   \textbf{4.  Stability Analysis of $f(T,B)$ cosmological models }
\end{center}
 \hspace{1cm}
 A differential equation is said to be an ordinary differential equation (ODE) if it contains only one independent variable, and one or more derivatives with respect to that variable. Let us consider an ordinary differential equation is of the form \cite{14}
   \begin{equation*}
     \dot{y}=\phi(y)
   \end{equation*}
    where $\dot{y}\equiv\frac{dy}{dt}$, $y=(y_{1},y_{2},y_{3},....,y_{n})\in\mathbb{R}^{n}$ and $\phi:\mathbb{R}^{n}\rightarrow \mathbb{R}^{n}$. When a system of ordinary differential equations does not explicitly depends on the independent variable, then the system is called an autonomous differential equation. In n dimensional space, the solutions $y_{1},y_{2},y_{3},....,y_{n}$ are the curves of this space. This space is known as phase space and the curves are known as phase trajectories. The phase space method is used to solve the time-dependent ordinary differential equations. To find the behaviour of dynamical system, we need to find the fixed points of this autonomous system. The fixed point of $\dot{y}=\phi(y)$ is a point $\tilde{y}\in\mathbb{R}^{n}$ such that $\phi(\tilde{y})=0$, i.e. a solution which does not change in time. Fixed points are classified as stable, unstable and saddle points that depends upon the stability. Let $\tilde{y}$ be a solution of $\dot{y}=\phi(y)$. Then $\tilde{y}$ is called stable fixed points if solutions start near to $\tilde{y}(t)$ at a given time and remains close to the fixed points for all later times\cite{15}. Mathematically, $\tilde{y}$ is stable if given $\epsilon>0$ $\exists$ a $\delta = \delta(\epsilon)>0$ such that if $h(t)$ be any other solution of the system $\dot{y}=\phi(y)$ satisfying $| \tilde{y}(t_{0}) - h(t_{0})|$ $<$ $\delta$, then the solution $h(t)$ exists $\forall$ $t$ $\geq t_{0}$ and it satisfies $|\tilde{y}(t) - h(t)|$ $< \epsilon$ $\forall$ $t>t_{0}$, $t_{0}$ $\in \mathbb{R}$ \cite{16}. The fixed point $\tilde{y}$ is said to be asymptotically stable if $\exists$ a constant $\delta>0$ such that if $|\tilde{y}(t_{0})-h(t_{0})|<\delta$ then $\lim_{t\rightarrow\infty}$$|\tilde{y}(t)-h(t)|=0$ \cite{17}. Suppose that $y=\tilde{y}$ be a fixed point of $ \dot{y}=\phi(y)$, $y\in \mathbb{R}^{n}$. Then $\tilde{y}$ is said to be hyperbolic fixed point if none of the eigenvalues of the Jacobian matrix at $y_{0}$, $J(y_{0})$ have zero real part, else the fixed point is called non-hyperbolic. For hyperbolic fixed points if all the eigenvalues of Jacobian matrix have positive real parts, then the fixed point is called repeller, unstable and trajectories are repelled from the fixed point. If all the eigenvalues of Jacobian matrix have negative real parts, then the point attract all nearby trajectories and is considered as stable fixed point and it is also known as attractor. If at least two eigenvalues of the Jacobian matrix have real parts with opposite signs, then the fixed point is called saddle point, which attracts trajectories in some directions but repels them along others directions.  Dynamical system analysis of the derived model can establish the stability of the derived cosmological model. Here, we assume two $f(T,B)$ cosmological models. The models are picked in order to obtain cosmological stable cases, in particular like late-time cosmological acceleration.\\
    Now we are living in a dark-energy- dominated and accelerated expansion of the Universe. Go back in time, the Universe used to be thick compared to today when matter dominated the Universe. Since in an expanding Universe radiation consumes faster than matter, dominating the world of radiation must be the first issue. Before to the radiation-dominated world, an accelerated expansion, called inflation, is trusted to have existed, which attached the beginning of the Universe Big Bang with the radiation era after inflation. Thus, the Universe has experienced two ages of acceleration; early time acceleration due to inflation and late time acceleration due to dark energy. Hence, any good cosmological model should contain at least part of the standard cosmological model which is shown as below \cite{18},
    \begin{center}
    Inflation  $\rightarrow$  Radiation  $\rightarrow$  Matter  $\rightarrow$  Accelerating Expansion.
    \end{center}
    To achieve the cosmological model from above relation, inflation points should be an unstable point while matter and radiation points should be a stable point for the Universe and accelerated phase should be an attractor. \\

 \textbf{4.1.{Power Law model in \textbf{$f(T,B)=f_{0}(B^{k}+T^{m})$} gravity}}

 Here we consider a power law model for $f(T,B)$ gravity which is given by
 \begin{equation}
   f(T,B)= f_{0}(B^{k}+T^{m})
 \end{equation}
where $f_{0}$, $k$ and $m$ are arbitrary constants. If $m<0$ then the Friedmann equations are affected in the late-time accelerating Universe, for $m>0$, this influence will be performed for the initial Universe \cite{19}. This analysis will be an effect when we include the boundary term $B$ on the collective evolution with $f(T,B)$ cosmology. Now, equation (19) obtained in terms of the dynamical variables as 
 \begin{equation}
   f_{T}= mf_{0}T^{m-1}
 \end{equation}
\begin{equation}
\dot{f_{T}}=\frac{m(m-1)f\dot{T}}{T^{2}}
 \end{equation}
\begin{equation}
f_{B}=\frac{kf}{B}
 \end{equation}
\begin{equation}
\frac{\dot{f_{T}}}{Hf_{B}}=-18\frac{m(m-1)}{k}(z-z^{2})
 \end{equation}
 Now, use Equation$(23)$ in Equations $(13)$-$(17)$, then the autonomous Equations can be written as :
 \begin{equation}
   x'=3-3x-6y-6z-6s+9\omega-x^{2}+3zs+3xz-12\frac{m(m-1)}{k}(z-z^{2})
 \end{equation}
 \begin{equation}
   y'=-6z+\lambda+6yz-3xy
 \end{equation}
 \begin{equation}
   z'=\lambda-6z^{2}
 \end{equation}
  \begin{equation}
   \omega'=-4\omega-3x\omega+6z\omega
 \end{equation}

  \begin{equation}
   s'=12\frac{m(m-1)}{k}(z-z^{2})-3xs
 \end{equation}
 Now we use this model to transform the equations as follows:
  \begin{equation}
  s=-\frac{2f_{T}}{3f_{B}}= -\frac{2m}{3k}(1+\frac{\dot{H}}{H^{2}})= -\frac{2m}{3k}(1-3z)
 \end{equation}
 This shows that $s$ is a function of $'z'$.
 \begin{equation}
   y=\frac{f}{18H^{2}f_{B}}= \frac{1}{k}(1+\frac{\dot{H}}{3H^{2}})=\frac{1}{k}(1-z)
 \end{equation}
 This shows that $y$ is also a function of $'z'$. With the help of equations $(31)$ and $(32)$, we reduced the autonomous system of differential equations to a $3$-D dynamical systems which are as follows:
 \begin{equation}
 x'= 3-3x-\frac{6}{k}(1-z)-6z+\frac{4m}{k}(1-3z)+9\omega-x^{2}-2\frac{m}{k}(1-3z)z+3xz-12\frac{m(m-1)}{k}(z-z^{2})
  \end{equation}
  \begin{equation}
    z'=\lambda-6z^{2}
  \end{equation}
  \begin{equation}
  \omega'=-4\omega-3x\omega+6z\omega
    \end{equation}
 To find the critical points, we find  $x'=0$, $z'=0$, and $\omega'=0$ to analyze the stability behaviors of the model. These systems have four critical points, which are shown in Table-$1$. For Model-$1$, effective EoS parameter $\omega_{eff}= -1-\frac{2\dot{H}}{3H^{2}}=-1+2z$ and deceleration parameter $q=-(1+\frac{\dot{H}}{H^{2}})=-(1-3z)$. So the critical points for these new autonomous system are in Table $1$.\\

\begin{tabular}{ |p{3cm}||p{3cm}|p{3cm}|p{3cm}|p{3cm}|  }
 \hline
 \multicolumn{4}{|c|}{\textbf{Table 1. Critical Points for Dynamical System}} \\
 \hline
 Critical Points&$ x$  &$ z$ &$\omega$&Exists When\\
 \hline
 A   & $(2\sqrt{\frac{\lambda}{6}}-\frac{4}{3})$    &$(\sqrt{\frac{\lambda}{6}})$& 0& $\lambda>0$\\
 \hline
 B &  $(2\sqrt{\frac{\lambda}{6}}-\frac{4}{3})$  &$(-\sqrt{\frac{\lambda}{6}})$& 0& $\lambda>0$\\
\hline
C &  $(-2\sqrt{\frac{\lambda}{6}}-\frac{4}{3})$ &$(\sqrt{\frac{\lambda}{6}})$& 0& $\lambda>0$\\
\hline
D & $(-2\sqrt{\frac{\lambda}{6}}-\frac{4}{3})$ &$(-\sqrt{\frac{\lambda}{6}})$& 0& $\lambda>0$\\
 \hline
\end{tabular}\\

For critical points which are shown in Table-$1$, eigenvalues are obtained from the Jacobian matrix which are as follows:

     \begin{equation}
    P_{1}=-\frac{1}{3}\pm\sqrt{\frac{\lambda}{6}}, \hspace{0.5cm} P_{2}= \pm12\sqrt{\frac{\lambda}{6}}, \hspace{0.5cm}  P_{3}=  0 .
    \end{equation}

\begin{tabular}{ |p{2cm}|p{1cm}|p{2cm}|p{3cm}|p{2cm}|p{2cm}|p{2cm}||  }
 \hline
 \multicolumn{6}{|c|}{\textbf{Table 2. Stability Conditions, deceleration parameter, density parameters}} \\
 \hline
 Critical Points&$ \Omega_{r}$  &$ \Omega_{m}$ &$q\newline(Exists for 0<\lambda<\frac{2}{3})$& $\omega_{eff}$ & $Stability$ & $Phase$\\
 \hline
 A   & $0$    &$1$& $-0.134$ & $-0.833$ & $unstable$ & $quintessence$  \\
 \hline
 B &  $0$  &$1$& $-1.866$& $-0.166$&$stable$ & $quintessence$\\
\hline
C &  $0$ &$1$& $-0.134$& $-0.833$&$unstable$ &$quintessence$\\
\hline
D & $0 $ &$1$& $-1.866$& $-0.166$&$stable$&$quintessence$\\
 \hline
\end{tabular}\\

     \begin{figure}
\includegraphics[height=2in]{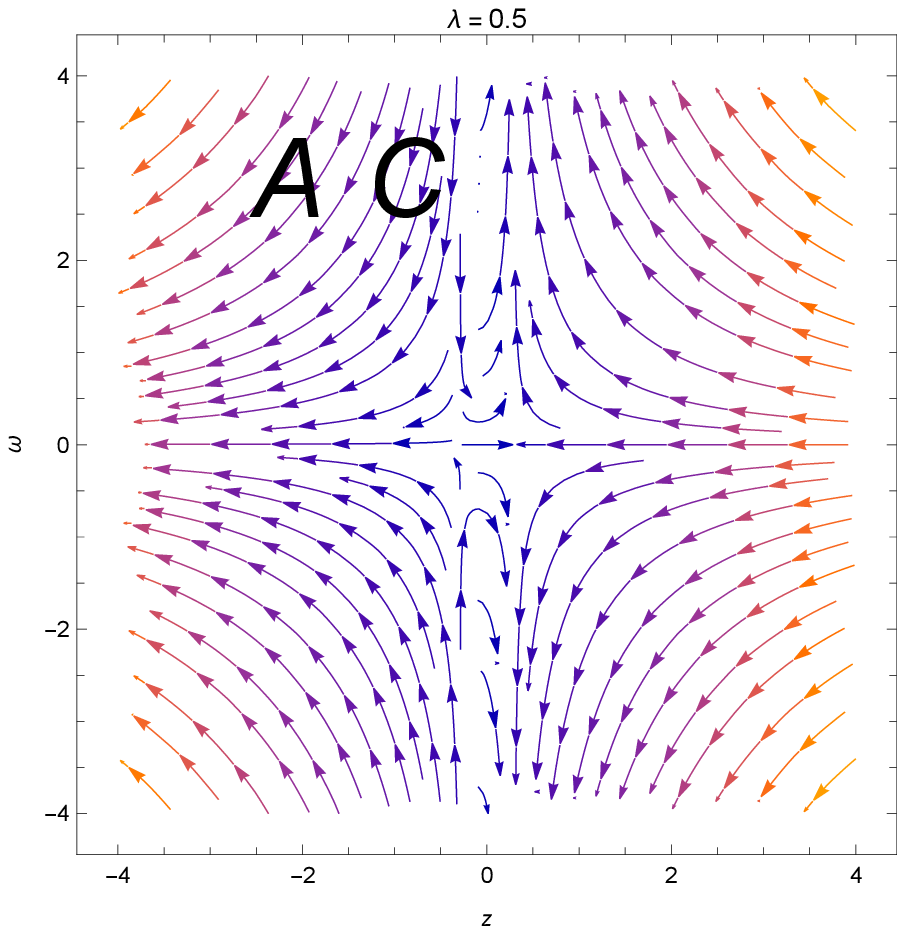}~~~~~~~~~
\includegraphics[height=2in]{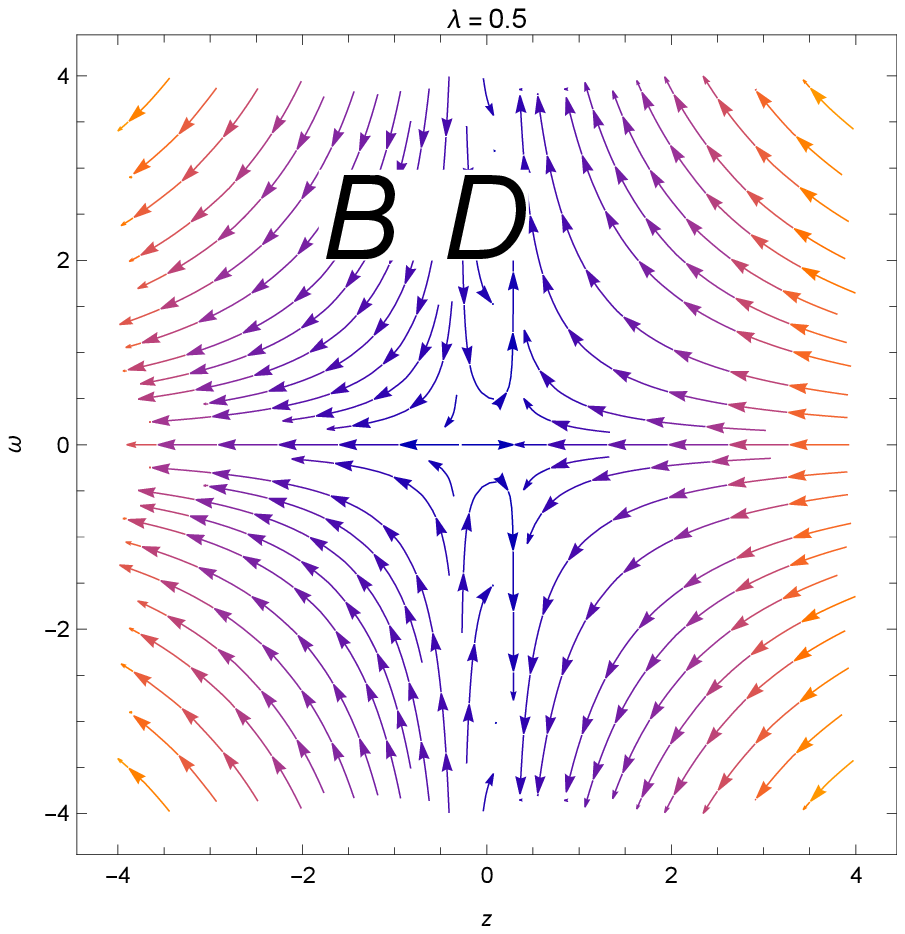}~~~~\\
{\bf{Figure. 1:} Phase portrait for Power Law Model. Here we plot the graph between $z$ and $w$. In the plot arrows represent the direction of the velocity fluid and the trajectories for this model. }

\hspace{1cm}\vspace{5mm}

\vspace{3mm}
\vspace{3mm}

\end{figure}

Figure-1 represents the trajectories of the phase space for different values of Table-$1$. From the figure the trajectories of the  critical points $A$ and $C$ move away from the fixed points. Hence these points ($A$,$C$) are unstable saddle. For critical points $B$ and $D$ the trajectories move towards from the fixed point. So these points ($B$, $D$) are stable point for $\lambda=0.5$.\\

Coley [25] have investigated that the dimension of the set of eigenvalues for non-hyperbolic critical points is one equal to the number of vanishing eigenvalues. Then the set of eigenvalues are normally hyperbolic and the critical points associated with it is stable but cannot be a global attractor. In our case, the set of eigenvalue is one and number of vanishing eigenvalue is also one. Hence we treat our obtained critical points as hyperbolic points. For the fixed points $A$ and $C$: the density parameters are $\Omega_{m}=1$, $\Omega_{r}=0$ and the EoS parameter $\omega_{eff}=-0.833$ which belongs to ($-1,0$) shows that the phase of the Universe is in quintessence epoch and $q<0$ shows that the Universe is accelerated for the region ($0<\lambda<\frac{2}{3}$). For these points ($A,C$), one eigenvalue($P_{1}$)is negative and another eigenvalue ($P_{2}$) is positive whereas one eigenvalue ($P_{3}$) is zero. It shows that the fixed points( $A,C$) are unstable saddle. For critical points $(B, D)$: the density parameters are $\Omega_{m}=1$, $\Omega_{r}=0$ and the EoS parameter $\omega_{eff}=-0.166$ which belongs to ($-1,0$) shows that the phase of the Universe is in quintessence and $q<0$ shows that the expansion of the Universe is accelerated for ($0<\lambda<\frac{2}{3}$). For these points ($B,D$) two eigenvalues ($P_{1},P_{2}$) are negative whereas one eigenvalue ($P_{3}$) is zero. Thus  the behavior of these points are stable. Stable and unstable fixed points for the different values of $\lambda$ are shown in Table-$2$. \\

\textbf{4.2.{Mixed Power Law model in $f(T,B)$=$f_{0}TB$ gravity }}

 In this form of the $f(T,B)$, the $f(T,B)$ model is given by \cite{18}
 \begin{equation}
   f(T,B)= f_{0}T B
 \end{equation}
 In this case,the model can be expressed as
   \begin{equation}
   f_{T}= 6 f_{0}(\dot{H}+T)
 \end{equation}
\begin{equation}
\dot{f_{T}}=6 f_{0}(\ddot{H}+12H\dot{H})
 \end{equation}
 \begin{equation}
 f_{B}= f_{0}T
 \end{equation}
 \begin{equation}
\frac{\dot{f_{T}}}{Hf_{B}}=-3\lambda-36z
 \end{equation}
 We use equations $(41-43)$ in $(17)$, we get
 \begin{equation}
x=\frac{\dot{f_{B}}}{3Hf_{B}}=\frac{\dot{T}}{3HT}=\frac{2\dot{H}}{H^{2}}=-2z .
 \end{equation}
 This imply that $x$ is a function of $z$.
 \begin{equation}
s=-\frac{2f_{T}}{3f_{B}}=12z-4 .
 \end{equation}
 This shows that $s$ is also a function of $z$. Using these cases, the autonomous system  of differential equations can be reduced as follows:
 \begin{equation}
y'=-6z+\lambda+12yz .
 \end{equation}
 \begin{equation}
z'=\lambda-6z^{2} .
 \end{equation}
 \begin{equation}
w'=-4w+12zw.
 \end{equation}
  To find the critical points,  we find  $x'=0$, $z'=0$ and $w'=0$  to analyze the stability behaviour of the model. These systems have $4$ critical points, which are shown as Table-$3$.  For Model-$2$, effective EoS parameter $\omega_{eff}= -1-\frac{2\dot{H}}{3H^{2}}=-1+2z$ and deceleration parameter $q=-(1+\frac{\dot{H}}{H^{2}})=-(1-3z)$. Critical Points for this model are in Table $3$:\\
 \newline
 \begin{tabular}{ |p{3cm}||p{3cm}|p{3cm}|p{3cm}|p{3cm}|  }
 \hline
 \multicolumn{4}{|c|}{\textbf{Table 3: Critical Points For this system of Equations}} \\
 \hline
 Critical Points&$ y$  &$ z$ &$w$&Exists When\\
 \hline
 A   & $\frac{6\sqrt{\frac{\lambda}{6}}-\lambda}{12\sqrt{\frac{\lambda}{6}}}$    &$(\sqrt{\frac{\lambda}{6}})$& 0& $\lambda>0$\\
 \hline
 B &  $\frac{6\sqrt{\frac{\lambda}{6}}-\lambda}{12\sqrt{\frac{\lambda}{6}}}$  &$(-\sqrt{\frac{\lambda}{6}})$& 0& $\lambda>0$\\
\hline
C &  $\frac{-6\sqrt{\frac{\lambda}{6}}-\lambda}{-12\sqrt{\frac{\lambda}{6}}}$ &$(\sqrt{\frac{\lambda}{6}})$& 0& $\lambda>0$\\
\hline
D & $\frac{-6\sqrt{\frac{\lambda}{6}}-\lambda}{-12\sqrt{\frac{\lambda}{6}}}$ &$(-\sqrt{\frac{\lambda}{6}})$& 0& $\lambda>0$\\
 \hline
\end{tabular}\\

For the critical points of Table-$3$, eigenvalues obtained from the Jacobian matrix are:
     \begin{equation}
    P_{1}=-4\pm12\sqrt{\frac{\lambda}{6}}, \hspace{0.5cm}  P_{2}=  \pm\sqrt{\lambda}\sqrt{6},\hspace{0.5cm}   P_{3}=  0 .
    \end{equation}\\

     \begin{figure}
\includegraphics[height=2in]{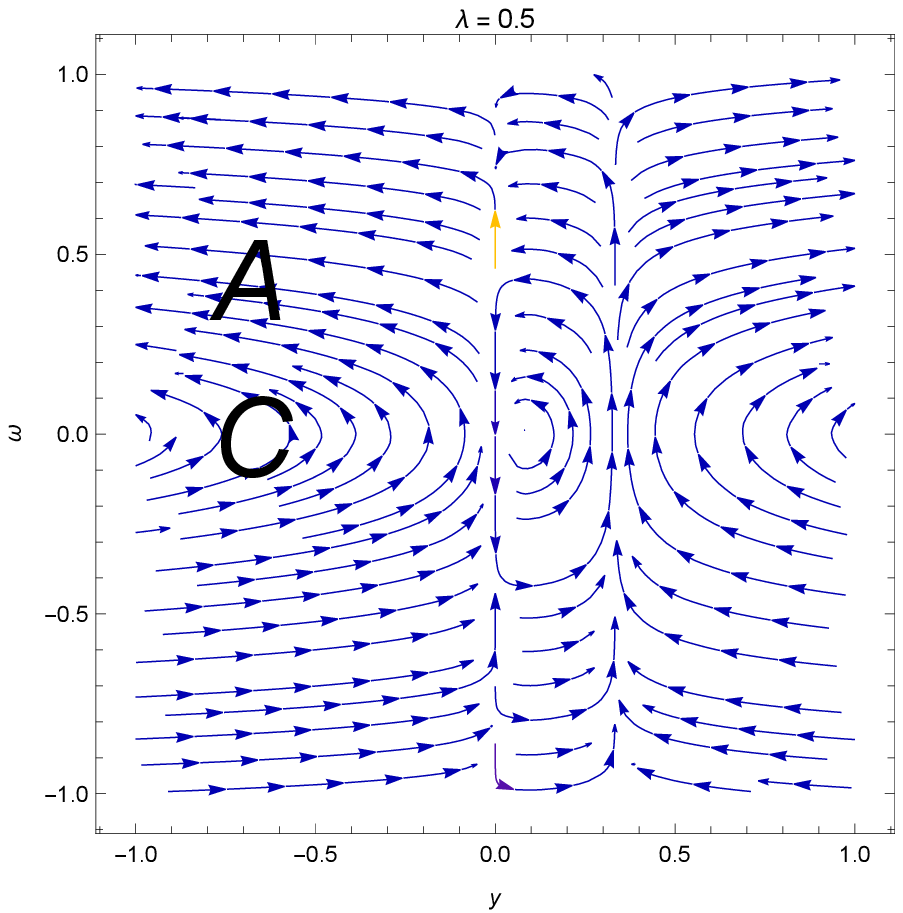}~~~~~~~~~
\includegraphics[height=2in]{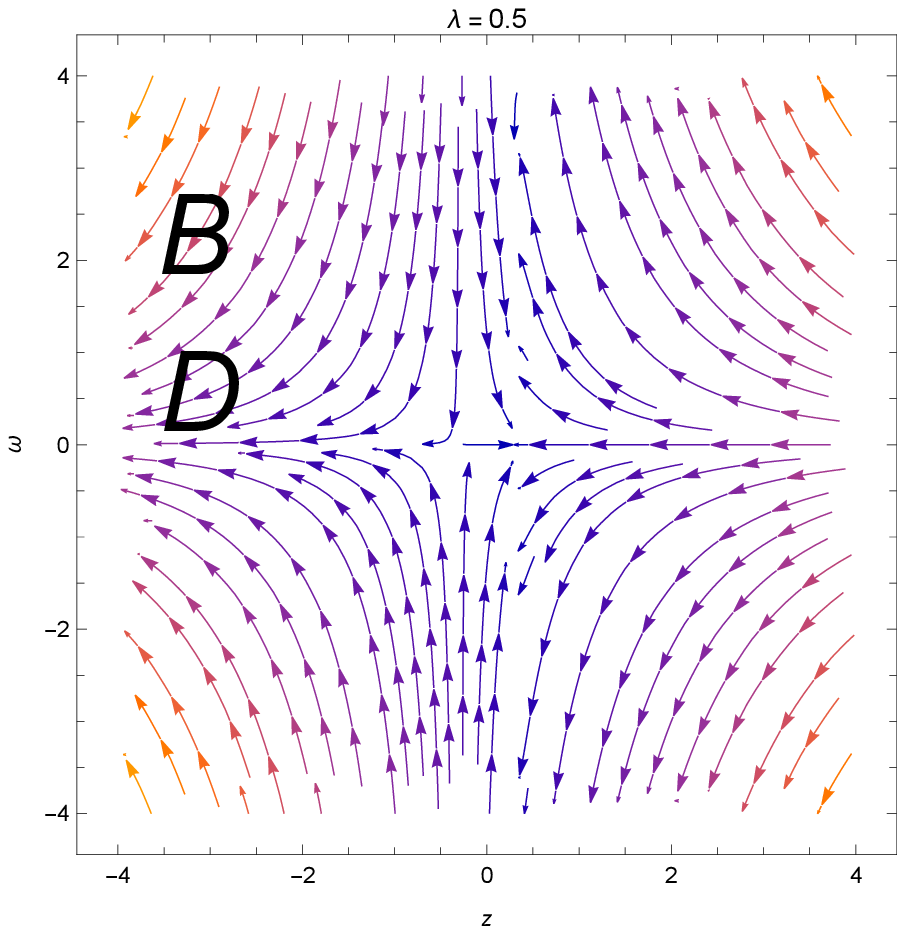}~~~~\\
{\bf{Figure. 2:} Phase portrait for Mixed Power Law Model. Here we plot the graph between  $y$ and $w$ and $z$ and $w$ . }

\hspace{1cm}\vspace{5mm}

\vspace{3mm}
\vspace{3mm}

\end{figure}
\begin{tabular}{ |p{2cm}||p{1cm}|p{2cm}|p{3cm}|p{2cm}|p{2cm}|p{2cm}|p{2cm}||  }
 \hline
 \multicolumn{6}{|c|}{\textbf{Table 4. Stability Conditions, deceleration parameter, density parameters}} \\
 \hline
 Critical Points&$ \Omega_{r}$  &$ \Omega_{m}\newline(for~~z=\frac{1}{3}, \lambda=\frac{2}{3})$ &$q\newline(for~~z=\frac{1}{3})$& $\omega_{eff}\newline(for~~z=\frac{1}{3})$ & $Stability$ &$Phase$\\
 \hline
 A   & $0$    &$1$& $0$ & $-\frac{1}{3}$ & $unstable$& $quintessence$ \\
 \hline
 B &  $0$  &$1$& $0$& $-\frac{1}{3}$&$stable$& $quintessence$\\
\hline
C &  $0$ &$1$& $0$& $-\frac{1}{3}$&$unstable$ &$quintessence$\\
\hline
D & $0 $ &$1$& $0$& $-\frac{1}{3}$&$stable$ &$quintessence$\\
 \hline
\end{tabular}\\

 Fig $2$, shows the phase portrait diagram for the dynamical system equations (44)-(46). From the Figure, the trajectories of the critical points $A$ and $C$ move away from the fixed points. Hence these points ($A$,$C$) are unstable saddle points. For critical points $A$ and $C$, the density parameters are $\Omega_{m}=1$, $\Omega_{r}=0$ and the EoS parameter $\omega_{eff}=-\frac{1}{3}$ shows that the phase of Universe is quintessence and $q=0$ shows that the Universe is marginal expansion for $z=\frac{1}{3}$ and $\lambda=\frac{2}{3}$. For the points ($A,C$) one eigenvalue $P_{1}$ is negative and another eigenvalue $P_{2}$ is positive whereas one eigenvalue ($P_{3}$) is zero. It shows that the fixed points ($A,C$) are an unstable saddle point. In second plot, trajectories for critical points $B$ and $D$ move towards to the fixed points. So the points ($B,D$) are stable point for $\lambda=\frac{2}{3}$. For these points ($B,D$), the density parameters are $\Omega_{m}=1$, $\Omega_{r}=0$ and the EoS parameter $\omega_{eff}=-\frac{1}{3}$ which shows that the Universe is in quintessence phase and $q=0$ shows that the Universe has marginal expansion for $z=\frac{1}{3}$ and $\lambda=\frac{2}{3}$. For these points eigenvalues are negative and zero. The negative and zero eigenvalues are demonstrates each other. The behaviour of these points are stable. Stable and unstable fixed points for the values of $\lambda$ are shown in Table-$4$. \\

 Now, we include cosmological constant in this system and consider that there is no interaction between any fluid particles. So, the field equations can be written as:
 \begin{equation}
  -3H^{2}(3f_{B}+2f_{T})+3H\dot{f_{B}}-3\dot{H}f_{B}+\frac{1}{2}f=k^{2}(\rho_{m}+\rho_{r}+ \rho_{\Lambda}) .
\end{equation}
\begin{equation}
   -(3H^{2}+\dot{H})(3f_{B}+2f_{T})-2H\dot{f_{T}}+\ddot{f_{B}}+\frac{1}{2}f=-k^{2}(\rho_{m}+\frac{4}{3}\rho_{r}).
\end{equation}
where $\rho_{\Lambda}$ is the cosmological density parameter. Now we introduce an extra dimensionless variable which gives,
\begin{equation}
  x=\frac{\dot{f_{B}}}{3Hf_{B}}, \hspace{0.5cm}   y=\frac{f}{18 H^{2}f_{B}}, \hspace{0.5cm}   z=-\frac{\dot{H}}{3H^{2}}, \hspace{0.5cm}   w=-\frac{\rho_{r}}{9H^{2}f_{B}}, \hspace{0.5cm}  s=-\frac{2f_{T}}{3f_{B}},  \hspace{0.5cm} r=-\frac{\rho_{m}}{9H^{2}f_{B}}
\end{equation}
The density parameters for matter and cosmological constants are,
\begin{equation}
 \Omega_{r}=-\frac{\rho_{r}}{9H^{2}f_{B}}=w \hspace{0.5cm} \Omega_{m}=-\frac{\rho_{m}}{9H^{2}f_{B}}=r,  \hspace{0.5cm}  \Omega_\Lambda=-\frac{\rho_{\Lambda}}{9H^{2}f_{B}}=1-x-y-z-s-w-r.
\end{equation}
From equation$(58)$ with the use of equation $(44)$, we derive the set of autonomous system of differential equations as follows:
\begin{equation}
  x'=3-3y-27z-3s+3r+12w+3zs+3xz+2\lambda-x^{2},
\end{equation}
\begin{equation}
  y'=-6z+\lambda+6yz-3xy,
\end{equation}
\begin{equation}
   z'=\lambda-6z^{2},
\end{equation}
\begin{equation}
  w'=-4w-3xw+6zw,
\end{equation}
\begin{equation}
  s'=-2\lambda+24z-3xs.
\end{equation}
\begin{equation}
  r'=-3r-3xr+6zr.
\end{equation}
From equation$(58)$, we can transform $x$ and $s$ as a variable of $z$, which is as follows:
\begin{equation}
x=\frac{\dot{f_{B}}}{3Hf_{B}}=\frac{\dot{T}}{3HT}=\frac{2\dot{H}}{H^{2}}=-2z .
 \end{equation}
 This shows that $x$ can be expressed as a function of $z$.
 \begin{equation}
s=-\frac{2f_{T}}{3f_{B}}=12z-4 .
 \end{equation}
 Using equations $(66)-(67)$, the autonomous system of differential equations would become,
 \begin{equation}
   y'=-6z+\lambda+12yz,
\end{equation}
\begin{equation}
  z'=\lambda-6z^{2},
\end{equation}
\begin{equation}
  w'=-4w+12zw=w(-4+12z),
\end{equation}
\begin{equation}
  r'=r(-3+12z),
\end{equation}
Here, we get four critical points. Critical points are shown in Table-$5$.\\
\newline
 \begin{tabular}{ |p{2cm}|p{3cm}|p{3cm}|p{3cm}|p{3cm}|p{3cm}| }
 \hline
 \multicolumn{6}{|c|}{\textbf{Table 5: Critical Points of this system of Equations}} \\
 \hline
 Critical Points&$ y$  &$ z$ &$w$ &r &Exists When\\
 \hline
 $A_{1}$  & $\frac{6\sqrt{\frac{\lambda}{6}}-\lambda}{12\sqrt{\frac{\lambda}{6}}}$    &$(\sqrt{\frac{\lambda}{6}})$& 0 & 0 &$\lambda>0$\\
 \hline
 $B_{1}$ &  $\frac{6\sqrt{\frac{\lambda}{6}}-\lambda}{12\sqrt{\frac{\lambda}{6}}}$  &$(-\sqrt{\frac{\lambda}{6}})$& 0&0 & $\lambda>0$\\
\hline
$C1$ &  $\frac{-6\sqrt{\frac{\lambda}{6}}-\lambda}{-12\sqrt{\frac{\lambda}{6}}}$ &$(\sqrt{\frac{\lambda}{6}})$& 0&0 &$\lambda>0$\\
\hline
$D_{1}$ & $\frac{-6\sqrt{\frac{\lambda}{6}}-\lambda}{-12\sqrt{\frac{\lambda}{6}}}$ &$(-\sqrt{\frac{\lambda}{6}})$& 0&0 &$\lambda>0$\\
 \hline
\end{tabular}\\

 For the critical points of Table-$5$, eigenvalues obtained from the Jacobian matrix are:

     \begin{equation}
    P_{1}=-4\pm12\sqrt{\frac{\lambda}{6}},  \hspace{0.5cm}  P_{2}=  \pm\sqrt{\lambda}\sqrt{6},   \hspace{0.5cm}  P_{3}=  0, \hspace{0.5cm}  P_{4}=-3\pm12\sqrt{\frac{\lambda}{6}}.
    \end{equation}\\

    The stability analysis and the values of density parameters of this system are shown in Table-6.\\
\newline
   \begin{tabular}{ |p{1.5cm}|p{1cm}|p{1cm}|p{2cm}|p{3cm}|p{2cm}|p{2cm}|p{2cm}||  }
 \hline
 \multicolumn{8}{|c|}{\textbf{Table 6. Stability Conditions, deceleration parameter, density parameters}} \\
 \hline
 Critical Points&$ \Omega_{r}$  &$ \Omega_{m}$ &$\Omega_{\Lambda}$ &$q\newline(for~~z=\frac{1}{3},\lambda=\frac{2}{3})$& $\omega_{eff}\newline(for~~z=\frac{1}{3})$ & $Stability$ &$Phase$\\
 \hline
$A_{1}$   & $0$  &$0$ &$1$& $0$ & $-\frac{1}{3}$ & $unstable$ &$quintessence$  \\
 \hline
$B_{1}$ &  $0$  & $0$ &$1$& $0$& $-\frac{1}{3}$&$stable$&$quintessence$\\
\hline
$C_{1}$ &  $0$  & $0$  &$1$& $0$& $-\frac{1}{3}$&$unstable$ &$quintessence$ \\
\hline
$D_{1}$ & $0 $  & $0$ &$1$& $0$& $-\frac{1}{3}$&$stable$&$quintessence$\\
 \hline
\end{tabular}\\

\begin{figure}
\includegraphics[height=2in]{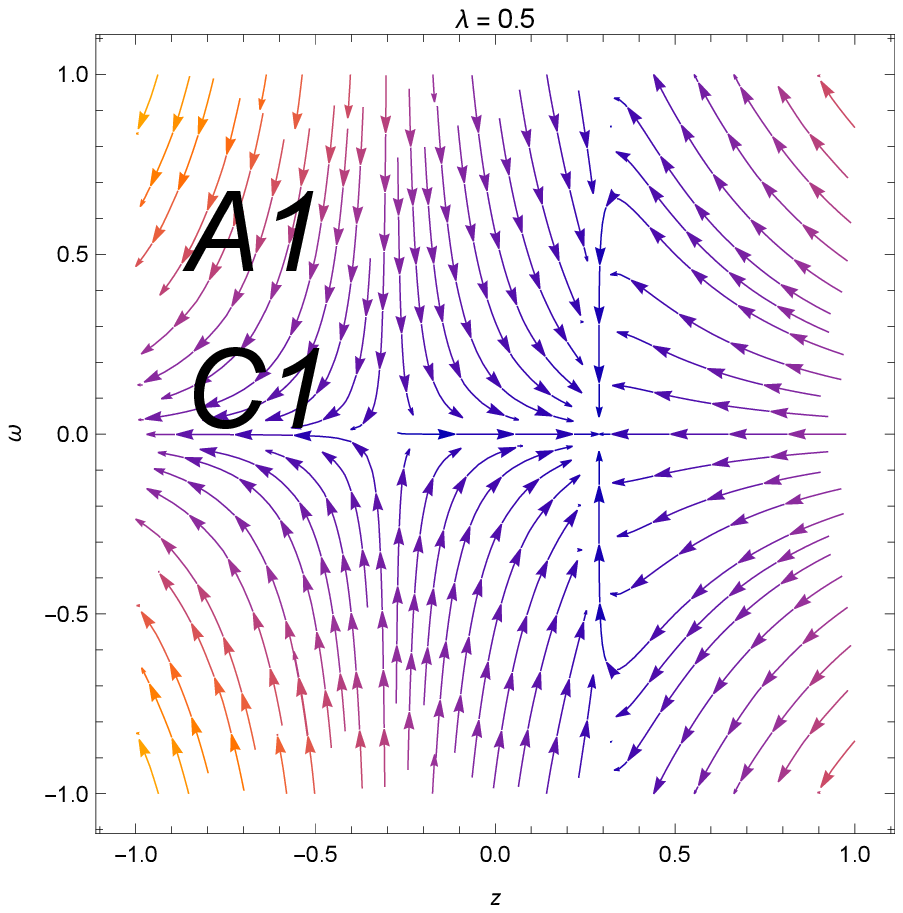}~~~~~~~~~
\includegraphics[height=2in]{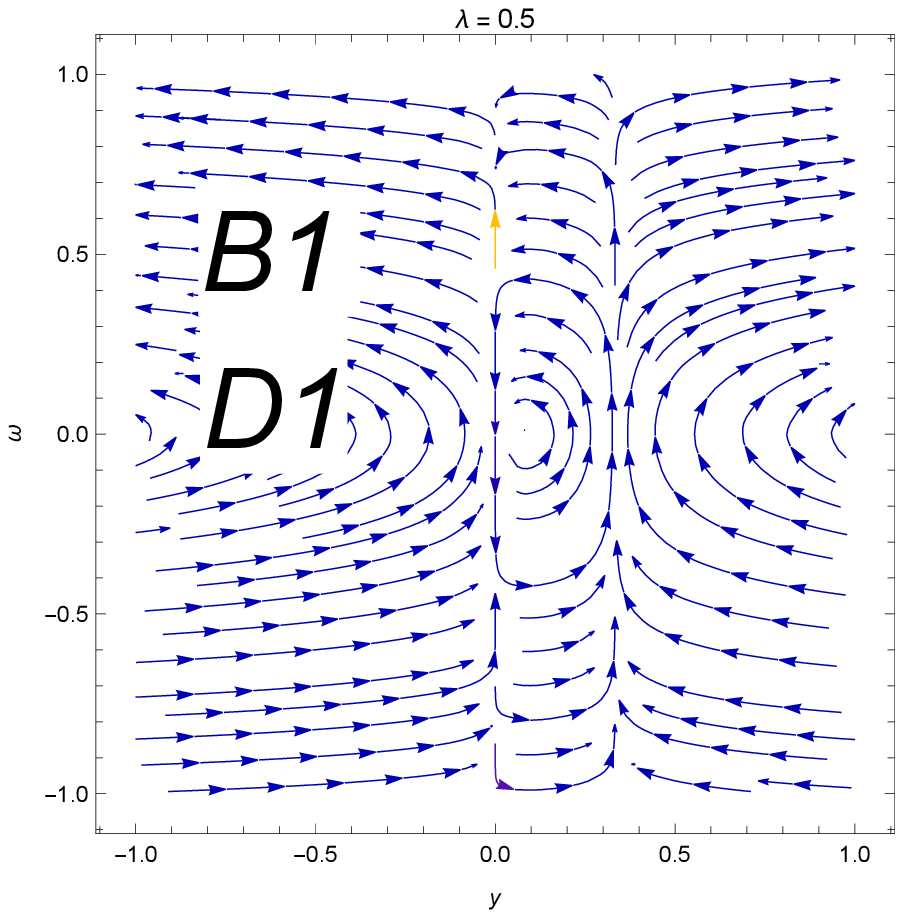}~~~~\\
{\bf{Figure. 3:} Phase portrait for Mixed Power Law Model. Here we plot the graph between $z$ and $w$ and $y$ and $w$. }
\hspace{1cm}\vspace{5mm}

\vspace{3mm}
\vspace{3mm}

\end{figure}

 Fig. $3$, shows that the phase portrait diagram for the dynamical system equations (60)-(63). From the figure the trajectories of the critical points $A_{1}$ and $C_{1}$ move away from the fixed points. Hence these points ($A_{1}$,$C_{1}$) are unstable saddle. We analyze that for fixed points $A_{1}$ and $C_{1}$, the density parameters are $\Omega_{m}=0$, $\Omega_{r}=0$ and $\Omega_{\Lambda}=1$ and the EoS parameter  $\omega_{eff}=-\frac{1}{3}$ shows that the Universe is in phase of quintessence and $q=0$ shows that the Universe shows marginal expansion for $\lambda=\frac{2}{3}$ and $z=\frac{1}{3}$. Here, we get phantom-like EoS ($\omega$$<$-1). For these points ($A_{1},C_{1}$), two eigenvalues ($P_{1},P_{4}$) are negative, one eigenvalue ($P_{3}$) is zero and one eigenvalue ($P_{2}$) is positive. Due to the presence of positive and negative eigenvalues, the behavior of critical points ($A_{1}$,$C_{1}$) are unstable saddle points. In second plot of Figure-$3$,  trajectories for critical Points $B_{1}$, $D_{1}$  move towards to the fixed points. So the points ($B_{1},D_{1}$) are stable point for $\lambda=\frac{2}{3}$. For these points the density parameters are  $\Omega_{m}=0$, $\Omega_{r}=0$ and $\Omega_{\Lambda}=1$ and the EoS parameter $\omega_{eff}=-\frac{1}{3}$ shows that the Universe is quintessence phase and $q=0$ shows that the Universe has marginal expansion for $\lambda=\frac{2}{3}$ and $z=\frac{1}{3}$. In these critical points, three eigenvalues ($P_{1},P_{2},P_{4}$) are negative and one eigenvalue ($P_{3}$) is zero. The negative and zero eigenvalues are demonstrates each other. The behavior of these points are stable. Stable and unstable fixed points for the values of $\lambda$ are shown in Table-$6$. \\

Now we look into the influences of interaction between the matter and dark energy. In present work, we consider only linear interaction. For interaction, the dimensionless variables have not been touched to each other. We take interaction $\bf Q$ between matter and dark energy. Then the continuity equations are written as \cite{20}
\begin{equation}
  \dot{\rho_{r}}+4H\rho_{r}=0,\hspace{0.4cm}  \dot{\rho_{m}}+3H\rho_{m}=Q, \hspace{0.4cm} \dot{\rho_{\Lambda}}=-Q.
\end{equation}
For this linear interaction $Q= H\rho_{tot}$ \cite{21}, The systems of autonomous differential equations for this linear interaction are as follows:
 \begin{equation}
  x'=3-3y-27z-3s+3r+12\omega+3zs+3xz+2\lambda-x^{2},
\end{equation}
\begin{equation}
  y'=-6z+\lambda+6yz-3xy,
\end{equation}
\begin{equation}
   z'=\lambda-6z^{2},
\end{equation}
\begin{equation}
  w'=-4w-3xw+6zw ,
\end{equation}
\begin{equation}
  s'=-2\lambda+24z-3xs .
\end{equation}
\begin{equation}
  r'=-1+x+y+z+2w+s-r-3xr+6zr.
\end{equation}
From equation $(58)$, we can transform $x$ and $s$ as a variable of $z$, which is as follows:
\begin{equation}
x=\frac{\dot{f_{B}}}{3Hf_{B}}=\frac{\dot{T}}{3HT}=\frac{2\dot{H}}{H^{2}}=-2z .
 \end{equation}
 This expression shows that $x$ can be expressed in terms of the variable $z$.
 \begin{equation}
s=-\frac{2f_{T}}{3f_{B}}=12z-4 .
 \end{equation}
 Using equations $(87)-(88)$, the set of autonomous differential equation can be reduced as:
 \begin{equation}
   y'=-6z+\lambda+12yz,
\end{equation}
\begin{equation}
  z'=\lambda-6z^{2},
\end{equation}
\begin{equation}
  w'=-4w+12zw=w(-4+12z),
\end{equation}
\begin{equation}
  r'=-5+y+11z+2w-r+12rz,
\end{equation}
For this system, we get $4$ critical points which are shown in Table-$7$ and  stability and acceleration analysis are discussed in Table-$8$.\\
\newline
 \begin{tabular}{ |p{2cm}|p{3cm}|p{3cm}|p{3cm}|p{3cm}|p{3cm}| }
 \hline
 \multicolumn{6}{|c|}{\textbf{Table 7: Critical Points of this system of Equations}} \\
 \hline
 Critical Points&$ y$  &$ z$ &$w$ & $r$ & Exists When\\
 \hline
 $A_{2}$  & $\frac{6\sqrt{\frac{\lambda}{6}}-\lambda}{12\sqrt{\frac{\lambda}{6}}}$ &$(\sqrt{\frac{\lambda}{6}})$& 0 & $\frac{21\lambda-54\sqrt{\frac{\lambda}{6}}}{12\sqrt{\frac{\lambda}{6}}-24\lambda}$ &$\lambda>0$\\
 \hline
 $B_{2}$ &  $\frac{6\sqrt{\frac{\lambda}{6}}-\lambda}{12\sqrt{\frac{\lambda}{6}}}$  &$(-\sqrt{\frac{\lambda}{6}})$& 0&$\frac{21\lambda+54\sqrt{\frac{\lambda}{6}}}{-12\sqrt{\frac{\lambda}{6}}-24\lambda}$& $\lambda>0$\\
\hline
$C_{2}$ &  $\frac{-6\sqrt{\frac{\lambda}{6}}-\lambda}{-12\sqrt{\frac{\lambda}{6}}}$ &$(\sqrt{\frac{\lambda}{6}})$& 0&$\frac{21\lambda-54\sqrt{\frac{\lambda}{6}}}{12\sqrt{\frac{\lambda}{6}}-24\lambda}$&$\lambda>0$\\
\hline
$D_{2}$ & $\frac{-6\sqrt{\frac{\lambda}{6}}-\lambda}{-12\sqrt{\frac{\lambda}{6}}}$&$(-\sqrt{\frac{\lambda}{6}})$&0&$\frac{21\lambda+54\sqrt{\frac{\lambda}{6}}}{-12\sqrt{\frac{\lambda}{6}}-24\lambda}$&$\lambda>0$\\
 \hline
\end{tabular}\\

   \begin{tabular}{ |p{1.5cm}|p{1cm}|p{1cm}|p{2cm}|p{3cm}|p{2cm}|p{2cm}|p{2cm}|  }
 \hline
 \multicolumn{8}{|c|}{\textbf{Table 8: Stability Conditions, deceleration parameter, density parameters}} \\
 \hline
 Critical Points&$ \Omega_{r}$  &$ \Omega_{m}$ &$\Omega_{\Lambda}$ &$q\newline(for~~z=\frac{1}{3}, \lambda=\frac{2}{3})$& $\omega_{eff}\newline(for~~z=\frac{1}{3})$ & $Stability$ &$Phase$\\
 \hline
 $A_{2}$   & $0$  &$1$ &$0$& $0$ & $-\frac{1}{3}$ & $unstable$ &$quintessence$ \\
 \hline
 $B_{2}$ &  $0$  & $1$ &$0$& $0$& $-\frac{1}{3}$&$stable$&$quintessence$\\
\hline
$C_{2}$ &  $0$  & $1$  &$0$& $0$& $-\frac{1}{3}$&$unstable$ &$quintessence$\\
\hline
$D_{2}$ & $0 $  & $1$ &$0$& $0$& $-\frac{1}{3}$&$stable$ &$quintessence$\\
 \hline
\end{tabular}\\

     \begin{figure}
\includegraphics[height=2in]{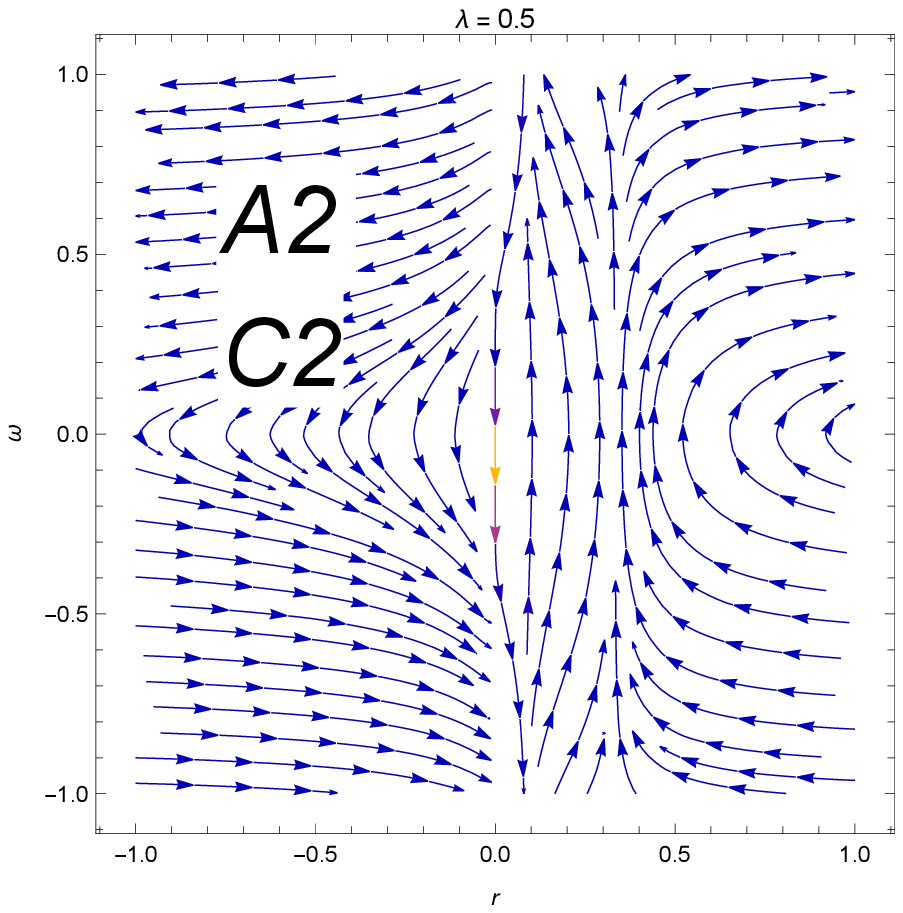}~~~~~~~~~
\includegraphics[height=2in]{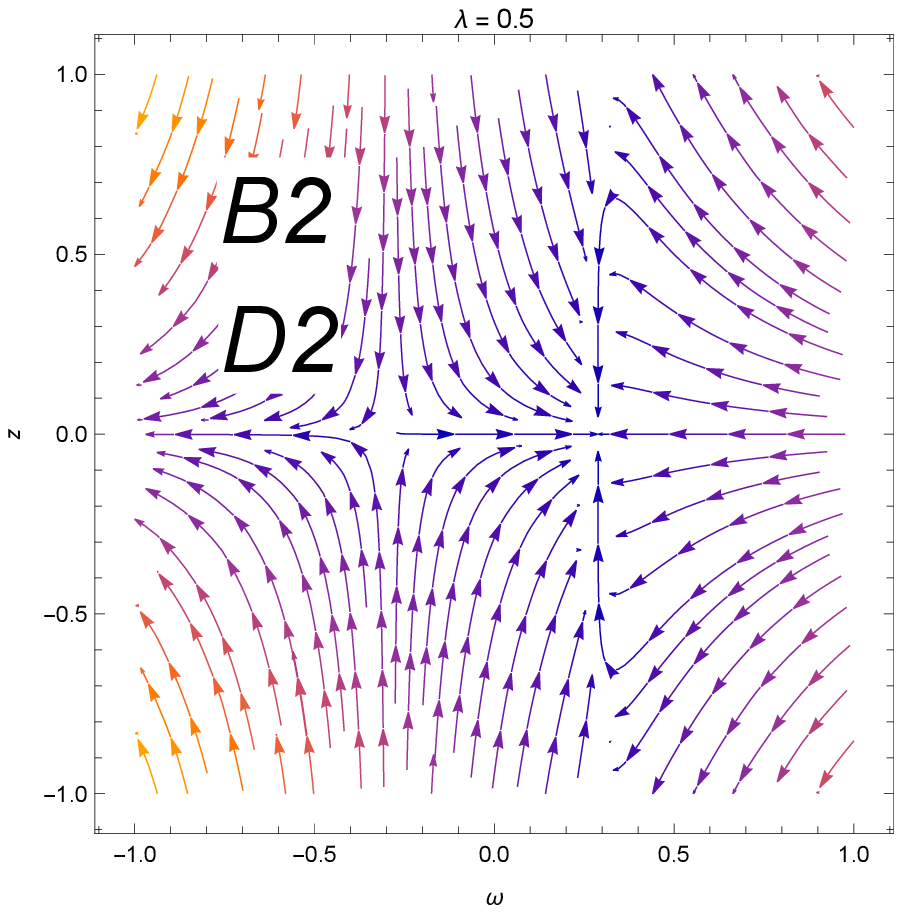}~~~~\\
{\bf{Figure. 4:} Phase portrait for Mixed Power Law Model for linear interaction. Here we plot the graph between $r$ and $w$ and $z$ and $w$.}

\hspace{1cm}\vspace{5mm}

\vspace{3mm}
\vspace{3mm}

\end{figure}
For the critical points of Table-$7$ , eigenvalues obtained from the Jacobian matrix are:
  \begin{equation}
    M_{1}=-4\pm12\sqrt{\frac{\lambda}{6}},  \hspace{0.5cm}  M_{2}= -1\pm12\sqrt{\frac{\lambda}{6}},   \hspace{0.5cm}  M_{3}=  \frac{-\sqrt{6\lambda}+\sqrt{102}\lambda}{2}, \hspace{0.5cm}  M_{4}= \frac{\sqrt{6\lambda}-\sqrt{102}\lambda}{2}.
    \end{equation}\\

 Fig. $4$, shows that the phase portrait diagram for the dynamical system equations (60)-(63). From the figure the trajectories of the critical points $A_{2}$ and $C_{2}$ move away from the fixed points.  Hence these points ($A_{2}$,$C_{2}$) are unstable saddle.  We analyze that for critical points $A_{2}$ and $C_{2}$ the density parameters are $\Omega_{m}=1$, $\Omega_{r}=0$ and $\Omega_{\Lambda}=0$ and the EoS parameter $\omega_{eff}=-\frac{1}{3}$ shows that the phase of the Universe is quintessence and $q=0$ shows that the Universe is marginal expansion for $\lambda=\frac{2}{3}$ and $z=\frac{1}{3}$. For these points ($A_{2},C_{2}$) three eigenvalues $M_{1}$, $M_{2}$ and $M_{4}$ are negative and one eigenvalue $M_{3}$ is positive.  Due to the presence of positive and negative eigenvalues, the behavior of critical points ($A_{2}$,$C_{2}$) are unstable saddle points. In second plot, the trajectories for critical points $B_{2}$ and $D_{2}$  move towards to the fixed points. So the points ($B_{2},D_{2}$) are stable point for $\lambda=\frac{2}{3}$. For critical Points $B_{2}$, $D_{2}$: the density parameters are $\Omega_{m}=1$, $\Omega_{r}=0$ and $\Omega_{\Lambda}=0$ and the EoS parameter $\omega_{eff}=-\frac{1}{3}$ shows that the phase of the Universe is quintessence and $q=0$ shows that the Universe is marginal expansion for $\lambda=\frac{2}{3}$ and $z=\frac{1}{3}$ . For these points ($B_{2},D_{2}$) all the eigenvalues $M_{1},M_{2},M_{3},M_{4}$ are negative real part. Hence the behaviour of these critical points are stable. Stable and unstable fixed points for the different value of $\lambda$ are shown in Table-$2$. \\

     \begin{figure}

\includegraphics[height=2in]{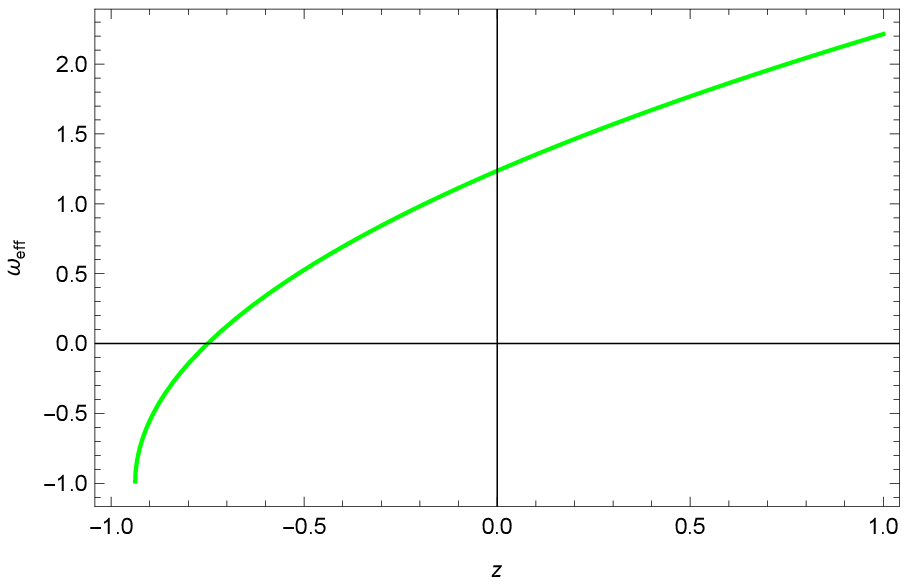}

{\bf{Figure. 5:}} Plot of EoS parameter ($\omega_{eff}$) versus time ($t$).

\hspace{1cm}\vspace{5mm}

\vspace{3mm}
\vspace{3mm}

\end{figure}

According to the definition of state parameter, if the value of EoS is exact $1$ then it represents the stiff fluid If the value is $0$ then it shows that the Universe is matter-dominated and when it is $1$, then the Universe is in radiation dominated phase. If the equation of state parameter lies between $-1$ and $0$ $i.e.$ $-1<\omega_{eff}<0$ then the Universe shows the quintessence phase
and when $\omega_{eff}=-1$ then it shows the cosmological constant $i.e.$ $\Lambda CDM$ model and when $\omega_{eff}<-1$ then phantom energy obtained. In our work, for first model we found the EoS parameter $\omega_{eff}=-0.833, -0.166$ for $(0<\lambda<\frac{2}{3})$ and for second model  $\omega_{eff}=-\frac{1}{3}$ for $z=\frac{1}{3}$ which shows that the Universe is in quintessence phase. Here we plot the graph of EoS parameter with respect to time(t) which is shown in Figure. $5$.\\

\begin{figure}

\includegraphics[height=2in]{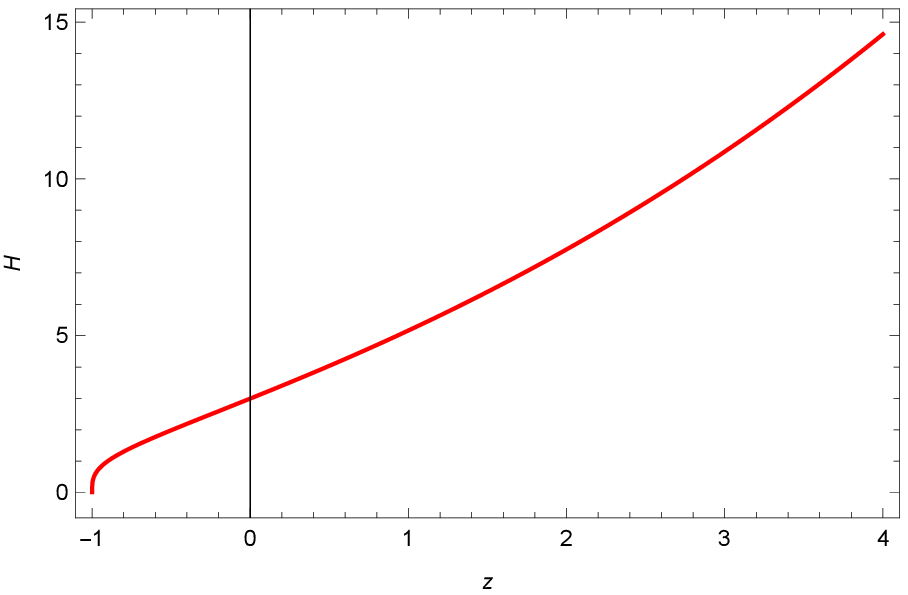}

{\bf{Figure. 6:}} Plot of Hubble parameter($H$) versus time ($t$).

\hspace{1cm}\vspace{5mm}

\vspace{3mm}
\vspace{3mm}

\end{figure}

For both models, we found that the scalar expansion and the Hubble parameter both are consistent for the expansion of Universe, which shows that the behaviour of Universe is expanding. The behaviour of Hubble parameter are shown in Figure-$6$. From Figure $7$, we notice that the behaviour of deceleration parameter is from negative to positive. In our work we get the value of $q(z)<0$ for first model and $q(z)=0$ for second model. For the value of $q(z)<0$ shows that the expansion Universe is accelerated and for $q(z)=0$ shows that the Universe is marginal expansion.\\

\begin{figure}

\includegraphics[height=2in]{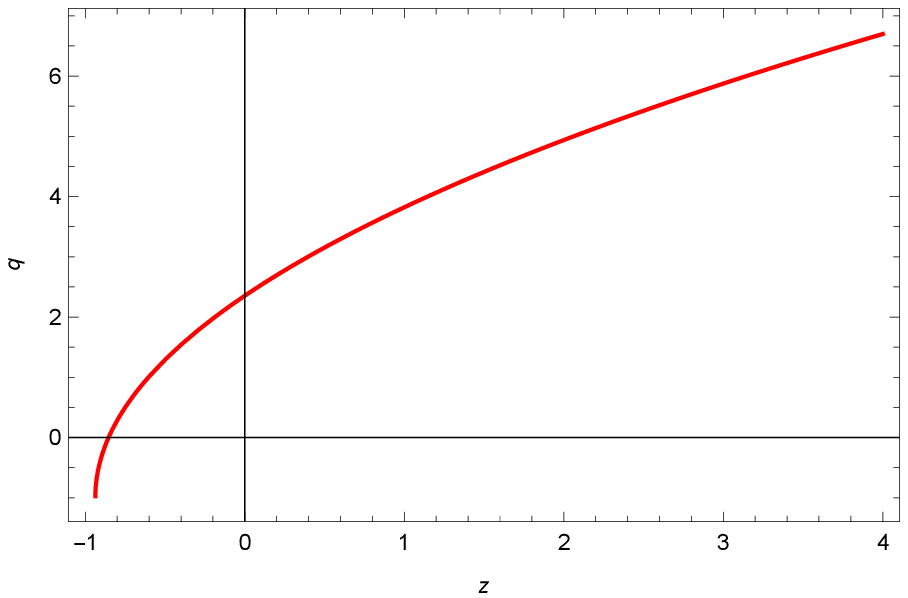}

{\bf{Figure. 7:}} Plot of deceleration parameter($q$) versus time ($t$).

\hspace{1cm}\vspace{5mm}

\vspace{3mm}
\vspace{3mm}

\end{figure}

\begin{center}
   \textbf{5. Conclusion }
\end{center}
 \hspace{1cm}
  In present days, modified gravity theories constructed on the basis of the accelerating expansion of Universe are one of the most important topics in research. From these theories, one of the theory is the new generalization of $ TG $ with torsion scalar(T) and boundary term(B) in the form of $f(T,B)$ \cite{22}, where $f(T,B)$=$f(-T+B)$=$f(R)$.   Here, we consider two different models of $f(T,B)$ gravity theory. The first model (power law model) is  $f(T,B)= f_{0}(B^{k}+T^{m})$, where $k$ and $m$ are arbitary real numbers. By introducing some dimensionless variables, we obtained exact cosmological solutions from the corresponding field equations using dynamical system analysis. We take the dynamical parameter $\lambda$ in terms of Hubble parameter in equation (18).  For this model, we get four critical points. In power law model, we obtain the effective EoS parameter $\omega_{eff}=-0.833$ for critical point $A$ and $C$ which is near to dark energy model. We get a unstable saddle point and for $B$ and $D$, we get the effective EoS parameter $\omega_{eff}=-0.166$ which is the quintessence epoch.  We get a stable node. All these values are in valid range with observational constants. The stability of the model has been observed when the EoS parameter is consistent with the $\Lambda$CDM model as $\omega_{eff}\sim -1$. These values of $\omega_{eff}$ show the accelerated expansion of the Universe. \\
  
  For model-$2$, $f(T,B)$ is taken to be taken to be $f(T,B)= f_{0}TB$. we get the effective EoS parameter ($\omega_{eff}=-\frac{1}{3}$) which is in quintessence epoch. Here, we get one unstable saddle node and one stable node. We consider the evolution of Universe in the presence of the interacting combination of matter, radiation and dark energy. We consider an linear interaction $Q=H\rho_{tot}$. We also analysis the stability and behaviour of critical points for this linear interaction. The main and interesting thing of these solutions are that they represent the quintessence of the Universe. We can write the deceleration parameter in terms of variable $z$ i.e. $q=3z-1$. This deceleration parameter can be written in terms of $\lambda$ i.e. $q=\pm3\sqrt{\frac{\lambda}{6}}-1$. We see that for first model $q(z)<0$, this represents that our Universe is in accelerating expansion and for second model we get $q(z)=0$ shows that the Universe is marginal expansion when $z=1/3$. From several cosmological project, the numerical value of EoS parameter as: $\omega_{eff}= -1.035^{+0.055}_{-0.059}$(Supernovae Cosmological Project), $\omega_{eff}= -1.073^{+0.090}_{-0.089}$(WMAP+CMB), $\omega_{eff}= -1.03\pm0.03$(Planck 2018). For both models we get the values of EoS parameter $ -1\leq\omega_{eff}<0$ which lies within these observational values. \cite{23}-\cite{24}. This work could be extended by considering different $f(T,B)$ cosmological model which satisfy all the conditions. In this paper the theoretical value of EoS parameter is important to explain the evolution history of the Universe. In both models, the behaviour of the dynamical parameters bounded by the scale factor and model parameters. The role of the parameter $\lambda$ has also an important role to better understand the dynamical behaviour of $f(T,B)$ gravity. Our models are simple but it gives better insights in the evolution of Universe.

 \end{document}